# How reproducible are methods to measure the dynamic viscoelastic properties of poroelastic media?

Paolo Bonfiglio[1], Francesco Pompoli[1], Kirill V. Horoshenkov[2], Mahmud Iskandar B Seth A Rahim[2], Luc Jaouen[3], Julia Rodenas[3], François-Xavier Bécot[3], Emmanuel Gourdon[4], Dirk Jaeger[5], Volker Kursch[5], Maurizio Tarello[5], Nicolaas Bernardus Roozen[6], Christ Glorieux[6], Fabrizio Ferrian[7], Pierre Leroy[8], Francesco Briatico Vangosa[9], Nicolas Dauchez[10] , Félix Foucart[10], Lei Lei[10], Kevin Carillo[11], Olivier Doutres[11], Franck Sgard[12], Raymond Panneton[13], Kévin Verdiere[13], Claudio Bertolini[14], Rolf Bär[14], Jean-Philippe Groby[15], Alan Geslain[16], Nicolas Poulain[17], Lucie Rouleau[18], Alain Guinault[19], Hamid Ahmadi[20], Charlie Forge[20]

[1]Department of Engineering (ENDIF), University of Ferrara, ITALY

[2]Department of Mechanical Engineering, University of Sheffield, Sheffield (UK)

[3]Matelys – Research Lab (France)

[4] ENTPE (France)

[5] Adler Pelzer Holding GmbH (Germany)

[6]Katholieke Universiteit Leuven (Belgium)

[7]STS-Acoustics (Italy)

[8]Saint-Gobain Isover (France)

[9] Dipartimento di Chimica, Materiali e Ingengeria Chimica “Giulio Natta”, Politecnico di Milano (Italy)

[10] Sorbonne Universités, Université de Technologie de Compiègne, Laboratoire Roberval (France)

[11] Department of Mechanical Engineering, École de Technologie Supérieure(Canada)

[12] IRSST (Canada)

[13] Université de Sherbrooke (Canada)

[14]Autoneum (Switzerland)

[15] Laboratoire d 'Acoustique de l'Université du Mans ( France)

[16] DRIVE EA1859, Univ. Bourgogne Franche Comté (France)

[17] Centre de Transfert de Technologie du Mans (France)

[18] LMSSC, Cnam (France)

[19] PIMM, ENSAM (France)

[20]Engineering & Design, Tun Abdul Razak Research Centre (TARRC (UK)

## Abstract
There is a considerable number of research publications on the acoustical properties of porous media with an elastic frame. A simple search through the Web of Science$^{TM}$ (last accessed 21 March 2018) suggests that there are at least 819 publications which deal with the acoustics of poroelastic media. A majority of these researches require accurate knowledge of the elastic properties over a broad frequency range. However, the accuracy of the measurement of the dynamic elastic properties of poroelastic media has been a contentious issue. The novelty of this paper is that it studies the reproducibility of some popular experimental methods which are used routinely to measure the key elastic properties such as the dynamic Young's modulus, loss factor and Poisson ratio of poroelastic media. In this paper, fourteen independent sets of laboratory measurements were performed on specimens of the same porous materials. The results from these measurements suggest that the reproducibility of this type of experimental method is poor. This work can be helpful to suggest improvements which can be developed to harmonize the way the elastic properties of poroelastic media are measured worldwide.

## I. INTRODUCTION

At the present time several analytical and numerical approaches are available to measure the vibro-acoustic performance of poroelastic materials used in noise and vibration control applications. Here we refer to those porous materials which frame can be treated as elastic, i.e. that has a finite value of the Young's modulus comparable to the bulk modulus of the air trapped in the material pores. Therefore, the average, or overall complex elastic moduli used in vibro-acoustic calculations is a combination of the elastic moduli of the material frame and air in the material pores. Commonly, these materials are characterised by the real part of the complex Young's modulus $E$ (hereafter storage modulus), loss factor $\eta$, and Poisson's ratio $\nu$.
The experimental determination of the elastic properties of viscoelastic solids as a function of frequency can be performed using different techniques. The choice of the appropriate measurement technique is influenced by the sample geometry, material damping factor and frequency range of interest. In some cases, the tested material specimen is preloaded with a static pressure in some others it is not. In some cases, the measurements are carried out over a broad range of temperatures whereas the frequency of excitation is unchanged, in others a range of excitation frequencies is applied at a given ambient temperature.

The strategy of this work is that there has been a number of inter-laboratory studies to understand the dispersion in the acoustical (surface impedance, sound absorption coefficient, characteristic impedance and complex wavenumber)[1-2] and related non-acoustical parameters (airflow resistivity, open porosity, tortuosity and characteristic lengths)[2-3] of porous media. However, the inter-laboratory studies on the elastic properties of porous media are much more scarce. The authors are aware of only one review of existing methods for determining elastic properties of materials was presented by Jaouen et *al*[4]. In this paper the authors compare different measurement techniques and apply them to melamine foam. To the best of knowledge of the authors there are no any other systematic studies which provide reliable experimental data and their dispersion in the elastic parameters of the same material specimens determined in several independent laboratories.

Therefore, the aim of this paper is to compare the results of some available methods which are used to measure the elastic properties of poro- and viscoelastic materials used in vibro-acoustic applications. Samples of the same materials are sent to a sufficiently large number of different laboratories in which a method used routinely to measure the elastic properties of porous media across a frequency and temperature range was applied. The data from these tested are then collated, analysed and presented in this paper. The novelty of this paper is that it provides a bespoke set of data which show the dispersion in the viscoelastic properties of the same porous media measured with different methods and in different laboratories around the world.

This paper is organised as follows: section II outlines the methodology; section III presents the results from individual laboratories and inter-laboratory data. Concluding remarks are made in section IV.

## II. METHODOLOGY

### A. Laboratories and tested materials

In this study fourteen acoustic research centres and private companies were involved. These are: University of Ferrara (Italy), Adler Pelzer Holding GmbH (Italy-Germany), STS-Acoustics (Italy), Polytechnic of Milan (Italy), University of Sheffield (UK)/TARRC (UK), Matelys Research Lab/ENTPE (France), Laboratoire d'Acoustique de l'Université du Maine (France)/LMSSC/Bourgogne, Cnam/PIMM (France), Laboratoire Roberval de l'Université de Technologie de Compiègne (France), Saint-Gobain Isover (France), Katholieke Universiteit Leuven (Belgium), Autoneum (Switzerland), IRSST/École de Technologie Supérieure (Canada) and Sherbrooke University (Canada). These centres were selected and contacted through a special call issued under the SAPEM[1] and DENORMS[2] networks. This enabled us to assemble a sufficiently large number of participants to cover a representative range of measurement techniques and to produce enough new data for the subsequent statistical analysis (see section II B). Some of the 20 partners were grouped in the following manner: the University of Sheffield worked with the Engineering & Design, Tun Abdul Razak Research Centre; the Université du Maine and University of Bourgogne teamed up with the CTTM; Matelys worked with ENTPE; CNAM teamed up with PIMM; and IRSST worked with École de Technologie Supérieure. The main reason behind this was to gain access to top of the range, state-of-the-art equipment for viscoelastic material testing and to bring in to this process a high level of expertise in porous media characterisation. These partnerships provided us with the opportunity to ensure a good consistency in sample preparation, testing and data interpretation. Specifically, this means that 14 sets of experiments were performed at the following 14 laboratories: University of Ferrara; ENTPE; Adler Pelzer Holding GmbH; Katholieke Universiteit Leuven; STS-Acoustics; Saint-Gobain Isover; Polytechnic of Milan; Laboratoire Roberval Centre de Recherches Royallieu; IRSST- École de Technologie Supérieure; Sherbrooke University; Autoneum; CTTM; LMSSC; and TARRC. This choice of laboratories was made to ensure that a range of measurement methods used by a majority of the research community and key material manufacturers to characterise the viscoelastic behaviour of porous media is well covered. Another criterion was the willingness of a particular laboratory to benchmark themselves publicly against other laboratories and to commit their time and resource to this set of voluntary experiments. The laboratory names were randomised to protect their identity, so each laboratory was assigned a unique id number between 1 and 14. Therefore, the laboratories are referred by only their id number in the following discussion.

Five different porous materials were investigated: reticulated foam, glass wool, porous felt, closed cell polyurethane foam and reconstituted porous rubber. These are denoted as materials

[1] Symposium on the Acoustics of Poro-Elastic Materials
[2] Action COST CA 15125

A, B, C, D and E, respectively. A description of tested materials is summarized in Table I. Figure 1 presents photographs of samples cut of the five materials.

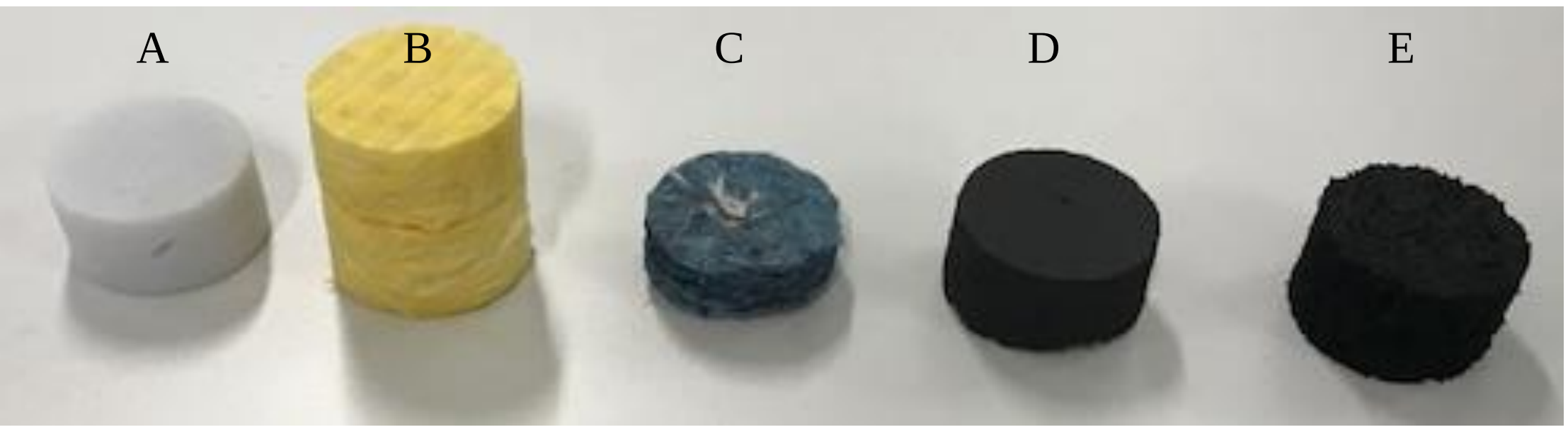

Figure 1 – Tested materials

Table I. Materials utilized in the inter-laboratory experiment.

| **Material** | **Description** | **Nominal thickness [mm]** | **Nominal Density [kg/m$^3$]** | **Airflow resistivity [Pa·s/m$^2$]** |
|---|---|---|---|---|
| A | Reticulated foam | 25 | 10 | ~ 10000 |
| B | Glass wool | 50 | 80 | ~ 70000 |
| C | Porous felt | 20 | 40 | ~ 80000 |
| D | Closed cell polyurethane foam | 25 | 48 | - |
| E | Reconstituted porous rubber | 25 | 240 | ~ 450000 |

Materials A-C are widely used for noise control. Material A represents a family of open cell foams. It is known to be one of the most homogenous and isotropic material and exhibits a relatively low dependency of its elastic properties on temperature and frequency[5]. Materials B and C represent the family of fibrous materials. They are anisotropic by structure, i.e. their depthwise elastic properties differ from those lengthwise and their stiffness increases with the static compression. Material D was chosen because it is a closed cell foam material and it shows a strong viscoelastic behavior[5]. Material E represents the family of consolidated granular material: it has a relatively high density, strong viscoelastic behaviour, and it is highly inhomogeneous due to rubber reconstitution process. This choice of materials covers a broad range of densities that is typical to those found in porous media used for noise control and vibration isolation. This material choice also reflects the fact that the elastic properties of porous media depend on the elastic parameters of the actual material frame and on the way the vibrating material frame interacts with the saturating air[6]. Sometimes these effects are separated by running two separate tests: (i) material sample is under the ambient atmospheric pressure; and (ii) material sample is in vacuum. The latter enables us to determine the elastic moduli of the material frame alone without the influence of the saturated air. However, this does not work with close cell foams because the air trapped in the close cells expands and alters significantly the overall elastic properties. Among the 14 laboratories only laboratory 11 carried out experimental tests on materials A-C and E in vacuum in addition to the ambient pressure test.

There are a number of effects which can lead to a noticeable dispersion in the elastic properties measured with different experimental techniques[4]. Firstly, it is the *inhomogeneity on a larger* *scale* due to the variability in the production process. As a result, some materials can exhibit differences in terms of their density and elastic properties. In this work, material slabs having the size of 40 cm x 100 cm were provided to each of the participating laboratory without any preliminary checks on their homogeneity. Secondly, the degree of anisotropy is typically different from one material to another. Thirdly, there may be some *effect of static preload and compression rate* which differ from test to test. It is common for viscoelastic materials such a porous media to show a dependency of the elastic properties on the initially applied static load or compression rate. In order to quantify some of these effects a detailed analysis was carried out on a particular measurement technique applied to same material specimen by different laboratories. This analysis is detailed in Section III.

### B. Measurement methods

Several measurement techniques of elastic properties were used by the participating laboratories. These measurement techniques can be divided in two distinct groups: (i) low frequency quasi-static methods; and (ii) dynamic methods. A further differentiation can be related to the type of the mechanical excitation applied to the sample. A majority of the 14 laboratories (except of laboratories 5, 10 and 14), measured longitudinal waves propagating along the thickness of the material sample. Laboratory 5 used in-plane flexural waves generated in the material slab. Laboratory 10 used the surface acoustics wave and laboratory 14 measured complex shear modulus by means of a torsional rheometer. A more detailed description of these measurement techniques is given in following sections.

*Quasi-static method*

The experimental set-up for a quasi-static compression test (hereafter indicated as QMA) consists of a sample sandwiched between two rigid plates. The lower plate is excited by an electrodynamics shaker and upper plate is rigidly fixed. According to the set-up a quasi-static compression test depicted in Figure 2a) three different quantities are measured in the frequency domain: (i) the vertical deformation ($D_1$) which is usually measured with accelerometer (2); (ii) the lateral deformation ($D_2$) which is usually measured with laser vibrometer (6); and the force transmitted through the tested material (F) measured with force transducer (3). Using these quantities it is possible to calculate the transfer function ($D_2/D_1$) and mechanical impedance ($F/D_1$) which are complex and frequency dependent for poroelastic media. Because the lower plate is excited, the dynamic force is applied upwards and the sample gets deformed in the longitudinal direction. In order to account for this effect (also known as “bulge effect”) a series of numerical simulations using finite element model is usually carried out. This enables us to determine the frequency dependent storage modulus $E$, the Poisson’s ratio $\nu$ and loss factor $\eta$. A more detailed description of the measurement technique can be found in ref. [8]. This methodology was adopted by laboratory 3.

Other laboratories used alternative approaches. Laboratories 2, 8 and 9 repeated the mechanical impedance test ($F/D_1$) on two samples of the same materials having different shape factors, $s=R/2L$, $R$ and $L$ being radius and thickness respectively as depicted in Figure 2b. It is strictly required that the two of samples are homogeneous and isotropic. As described in refs. [9] and [10], a series of preliminary finite element simulations can be carried out to account for the "bulge effect" through polynomial relations to determine Young's modulus, Poisson ratio and loss factor. All these laboratories set the Poisson's ratio to 0 for materials B and C, which was a usual choice for highly porous fibrous materials. Laboratory 6 utilised a similar approach and measured the mechanical impedance ($F/D_1$) of a single sample of each material assuming a known value for Poisson's ratio based on microstructure consideration[9,11]. In these particular tests two hypotheses were given for the Poisson's ratio that was set to 0.33 or 0.45 for materials A, D and E. For materials B and C, the Poisson's ratio was set to 0 as commonly accepted for such materials. Laboratory 10 determined storage modulus and loss factor directly from longitudinal stiffness tests through measuring the mechanical impedance ($F/D_1$) and by setting Poisson's ratio to zero[8,12].

*Resonant method/Transmissibility based method*

The original method is described in detail in ref. [13]. The bottom of a rectangular specimen is loaded with a mass. The top surface of the specimen is attached to a rigid rectangular plate which is excited with a shaker. According to the set-up shown in Figure 2c this technique is based on the measurement of the amplitude of the transmissibility function that is the ratio between top and bottom plate accelerations determined in a broad frequency range. The resonance frequency and quality factor can then be determined from this frequency dependent transmissibility function and related unambiguously to the Young's modulus and loss factor of the material specimen. In this test the Poisson's ratio cannot be measured and it is usually set to zero. This experimental methodology was adopted by laboratories 1 and 4. Laboratory 7 and 11 tested samples with different shape factors and made use of polynomial relationships (approach similar to that described in ref. [9]) in order to estimate the Poisson's ratio. This approach is depicted schematically in Figure 2d.

*Dynamic mechanical analysis and time-temperature superposition principle*

Dynamical mechanical analysis (DMA) is an experimental technique commonly used to study the frequency and temperature dependence of the elastic properties of viscoelastic materials. In order to determine the mechanical response of a viscoelastic material (e.g. polymers or polymer based composites) to a sinusoidal strain/stress over an extended range of frequencies, it is possible to perform tests over a limited range of frequencies but over an extended temperature range. The "time temperature equivalence (TTS)[14-16] can then be exploited to generate the so called "master curve" from which the elastic properties of this material specimen (e.g. the Young's modulus and loss factor) can be determined at a given temperature but over an extended range of frequencies.

Laboratories 12 - 14 used a standard dynamic mechanical analyser which was able to measure the Young's modulus and loss factor. Laboratory 13 applied the TTS principle to material D excited in compression to estimate the Young's modulus and loss factor over a much more extended frequency range than that achieved by laboratory 12. Laboratory 14 utilised a similar approach but with the sample excited in torsion over a limited frequency range and applied the TTS principle to materials A and D (Figure 2e)[17] to extend this range considerably. These laboratories did not measure the Poisson's ratio and assumed it was equal to zero.

*Lamb wave propagation and surface acoustic wave method*

The method adopted by laboratory 5 is explained schematically in Figure 2f. A slab of porous material was fixed on one side and its other edges were left free to vibrate. The material was excited using an electromagnetic shaker at one point and normal displacement was measured at different distances from the source using a laser vibrometer with a fixed spatial step of 5 mm. The geometrical dispersion of propagating Lamb waves was accounted for with a model which enabled this laboratory to invert the elastic properties of the porous material slab material[18,19].

Laboratory 10 performed a measurement on one sample of material A using a spatial Laplace Transform for complex wavenumber approach[20] experiment which is illustrated in Fig 2g. The bottom of the material slab and its right-hand edge were glued to a rigid hard surface. The left-hand edge was excited with a shaker and the normal displacement of its top surface was measured using a laser vibrometer over a 60 cm span with a fixed spatial step of 0.5 mm. The method was applied to determine the real and imaginary parts of the wavenumber for the guided elastic wave excited in the porous slab and then, using the dispersion relationship for Rayleigh waves, to estimate the values of the complex Young's modulus and Poisson's ratio.

*Transfer function/ transfer matrix method*

In these experiments the tested material was assumed to be homogeneous and isotropic. The material sample was mounted on a support plate which was excited by an electromagnetic shaker as it is shown in Figure 2h. Using a logarithmic sine sweep as the excitation signal, the acceleration of the bottom plate was measured using an accelerometer, and the velocity at the top surface of the sample was determined using a laser vibrometer as shown in Figure 2h. For a harmonic excitation and assumed value of the Poisson's ratio it was possible to calculate the complex Young's modulus through the plane wave transfer matrix approach for wave propagation in an elastic solid using the measured downstream-upstream velocity transfer function across a test sample. A detailed description of the measurement technique is given in ref. [5].

In this project all the participating laboratories measured the complex Young's modulus for all the materials. Not all of the 14 laboratories had the equipment and expertise to measure the Poisson's ratio of porous media. In fact, measurement of the Poisson's ratio of porous media remains a challenging and the quality of the data obtained from these tests is often controversial (e.g. ref. [4]). Therefore, the Poisson's ratio data only came out of those laboratories who had confidence in their data and techniques used to obtain them. Table II provides a list of the laboratories who measured the Poisson's ratio. Tables III and IV give a summary of the measurement setups and procedures used in the reported experiments.

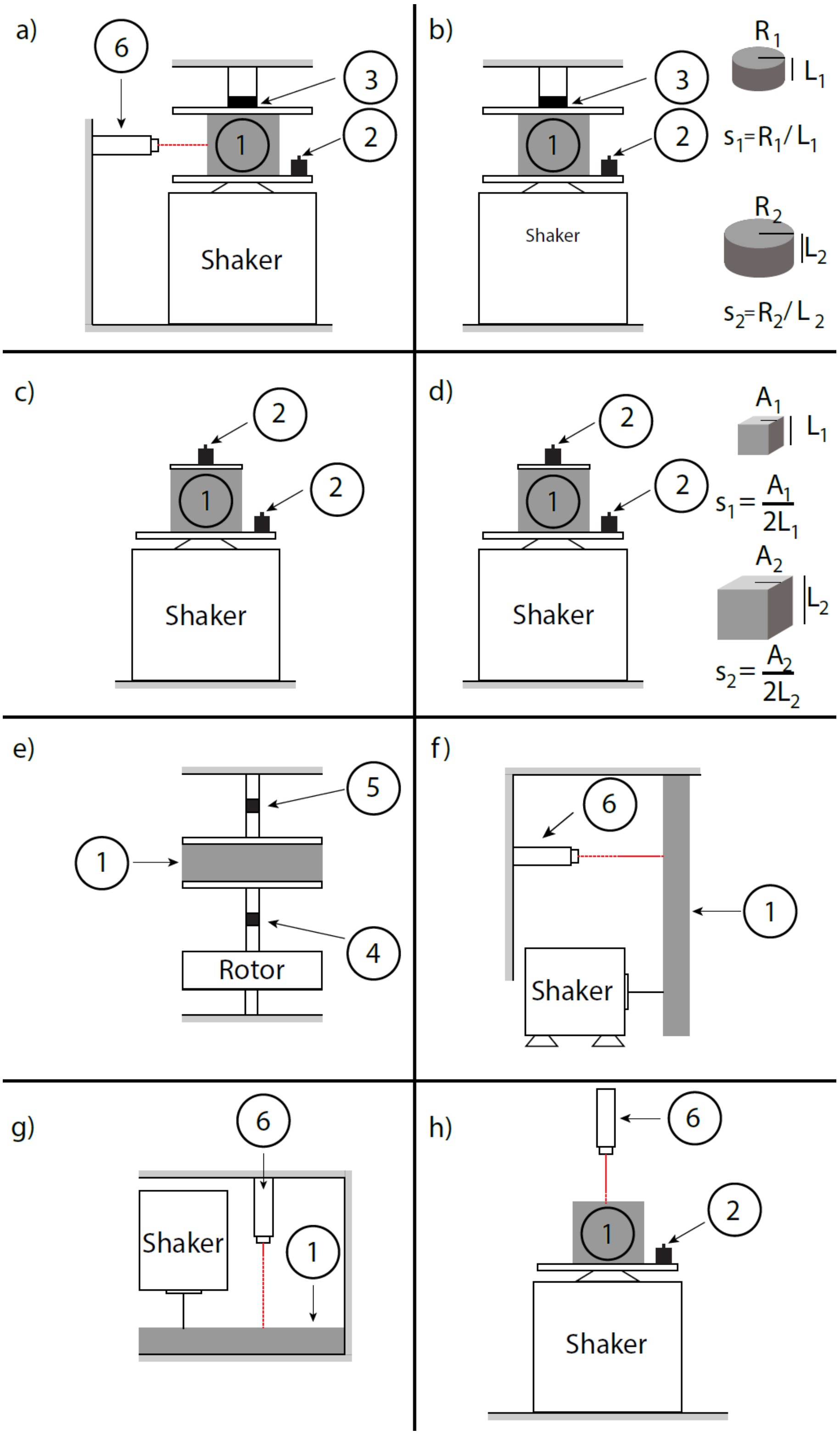

Figure 2 – Basic measurement setups for: a) and b) quasi-static uniaxial compression methods, c) and d) resonant methods, e) dynamic torsional method, f) Lamb wave propagation method, g) Surface acoustic wave method, h) transfer function/transfer matrix method. 1-sample; 2-accelerometer; 3-force transducer;4-torque transducer; 5-angular displacement transducer; (6) laser vibrometer.

Table II. Summary of Poisson's ratio measurement (●: measured, empty: not measured, numerical: fixed value). The letters A and B suggest that the same laboratory used two different measurement methods.

| | Partner | | | | | | | | | | | | | | | |
|---|---|---|---|---|---|---|---|---|---|---|---|---|---|---|---|---|
| **Material** | 1 | 2 | 3 | 3B | 4 | 5 | 6 | 7 | 8 | 9 | 10 | 10B | 11 | 12 | 13 | 14 |
| A | 0 | ● | ● | From method 3 | 0 | 0 | 0.33 or 0.45 | ● | ● | ● | 0 | ● | ● | 0 | 0 | 0 |
| B | 0 | ● | ● | From method 3 | 0 | 0 | 0 | ● | 0 | 0 | 0 | | 0 | 0 | 0 | |
| C | 0 | 0 | ● | From method 3 | 0 | 0 | 0 | ● | 0 | 0 | 0 | | 0 | 0 | 0 | |
| D | 0 | ● | ● | From method 3 | 0 | 0.33 | 0.33 or 0.45 | ● | ● | ● | 0 | | ● | 0 | 0 | 0 |
| E | 0 | ● | ● | From method 3 | 0 | 0 | 0.33 or 0.45 | ● | ● | ● | 0 | | ● | 0 | 0 | |

Table III. Summary of measurement techniques used by the 14 participating laboratories (R: radius, LS: lateral side). The letters A and B suggest that the same laboratory used two different measurement methods.

| **Laboratory** | ***Method*** | ***Measurement set-up*** | ***Frequency range*** | ***# of tested samples for each materials*** | ***Size of specimen [mm]*** | ***Reference*** |
|---|---|---|---|---|---|---|
| 1 | Resonant | Fig. 2c | Value at resonance frequency | 5 | 50 (LS) | Not declared |
| 2 | QMA | Fig. 2b | 10-60 Hz step 10 Hz | 5 | 20 and 50 (R) | 9,10 |
| 3 | QMA | Fig. 2a | 20-45 Hz | 5 | 22. 5 (R) | 7,8 |
| 3B | Transfer Function/Transfer Matrix | Fig. 2h | 60-1000 Hz (60-300 Hz for material C) step 0.5 Hz | 1 | 22. 5 (R) | 5 |
| 4 | Resonant | Fig. 2c | Value at resonance frequency | 5 | 49 (R) | Internal measurement protocol |
| 5 | Lamb wave | Fig. 2f | 100-1000 Hz | 1 | 40 x 100 $cm^2$ | 18,19 |
| 6 | QMA | Fig. 2a | 20-120 Hz step 10 Hz | 5 | 22.25 (R) | 9,11,24 |
| 7 | Resonant | Fig. 2d | Value at resonance frequency | 5 | 50-100 (LS) | 6,9,13 |
| 8 | QMA | Fig. 2b | 20-40 Hz step 5 Hz | 5 | 44.4 and 29 (R) | 9,10 |
| 9 | QMA | Fig. 2b | 10-60 Hz step 10 Hz | 5 | 15 and 22.25 (R) | 9,10 |
| 10 | QMA | Fig. 2a | 10-40-70 and 100 Hz | 3 | 22.25 (R) | 8,12 |
| 10 B | SAW | Fig. 2g | Single value that fit the data in the frequency range of 200-4000 Hz | 1 | 40 x 100 $cm^2$ | 20 |
| 11 | Resonant | Fig. 2d | 40-500 Hz step 10 Hz | 5 | 50 (R) and circular annular | 9 |
| 12 | DMA | Fig. 2a | 0.1-100 Hz -log step | 1 | 14.5-17.5 (R) | 14,15,16 |
| 13 | DMA+TTS | Fig. 2a | 0.1-10 Hz. (0.1-5.4e8 Hz for material D) - log step | 1 | 15 (LS) | 14,15 |
| 14 | DMA+TTS | Fig. 2e | 0.1 – 5e5 Hz - log step | 1 sample of materials A and D | 12 (R) | 17 |

Table IV. Description of measurement procedures used by the 14 participating laboratories.

| Laboratory | *Excitation signal* | *Calibration procedure* | *Static load / compression rate/ imposed dynamic amplitude* | *T [°C]* | *Method of support the samples* |
|---|---|---|---|---|---|
| 1 | Random signal | Accelerometer amplitude calibration | 50.2 g by top plate | 22 | The sample is bonded on bottom and top plates |
| 2 | Pure tones | Force sensor and accelerometer couple checked by measuring the stiffness of a reference spring. | Compression rate for foams is fixed to value which guarantees constant stiffness, for fibrous materials is fixed to 1.7%<br>Dynamic amplitude: fixed to 5e-6m. | 25 | Contact |
| 3 | Sine sweep | Force sensor and accelerometer couple checked by measuring the stiffness of a reference spring. | Measurement at different static load and extrapolation at zero static force. | 23 | Glue between sample and plates |
| 3B | Sine sweep | A calibration function in frequency domain is determined by measuring the response of the bottom plate without sample. | No static load is applied. | 23 | Glue between sample and plates |
| 4 | Chirp | The amplitude of the transfer function between the accelerometers is checked to be less than 1,01 | 134.4 g or 547.3 g, depending on sample stiffness by top plate. The dynamic amplitude in not fixed. | 21 | Glue between sample and plates |
| 5 | Sine sweep | No calibration is required | No static load/compression rate is applied. | 22 | Material is freely suspended and clamped at top edge |
| 6 | Pure tones | Force sensor and accelerometer couple checked by measuring the stiffness of a reference spring. | Compression rate fixed to 0% | 18-21 | The sample is glued on bottom and top plates |
| 7 | Pseudo Random Noise | Accelerometers are calibrated measuring the same FRF of the base plate | Between 82 gr and 192 gr depending on material stiffness and surface aspect | 18 | Two sided bonded tape between sample and plates |
| 8 | Pure tones | Force sensor and accelerometer couple checked by measuring the stiffness of a reference spring. | Compression rate:<br>- for foams is fixed to value which guarantees constant stiffness;<br>- for fibrous materials is fixed to 1 – 6 % | 20 | Sand paper between sample and plates |
| 9 | Pure tones | Calibration from manufacturer | Compression rate fixed to 1.7 – 3 % | 23 | Sand paper between sample and plates |
| 10 | Pure tones | Force sensor and accelerometer couple checked by measuring the stiffness of a reference spring. | Compression rate fixed to 3% | 22 | Contact |
| 10B | Pure tones | No calibration is applied | No static load/compression rate is applied. | 22 | The sample is glued on a rigid floor |
| 11 | White noise | The transmissibility function between the accelerometers is checked to be 0dB +/- 0.1dB and +/- 3deg for phase up to 1kHz | Mass load chosen in order to have a compression rate lower than 2%. For material C compression rate was fixed to 5 %. | 23 | Two sided bonded tape between sample and plates |
| 12 | Sweep sine | No calibration is applied | Compression rate: 5 % | 23 | Contact |
| 13 | Sweep sine | Force transducer calibrated using a precision weight | Static pre-strain: 5 % for materials A, B, D and E. 30% for material C. Strain amplitude 0.1% | 23 (add. temperatures in order to apply TTS). | Contact |
| 14 | Sweep sine | No calibration is applied | No static preload | 20 (add. temperatures in order to apply TTS) | Two sided bonded tape between sample and plates |

### C. Error analysis

A key aim of the interlaboratory test was to determine the repeatability and reproducibility variances of the test methods adopted by the partners. The statistical procedures prescribed in the ISO 5725-1 and 5725-2 standards[21,22] were used for this purpose. Although the ISO 5725 series standards refer to the same measurement method, it is believed that they can give strong indication about the consistency of measurement data from different laboratories using different measurement techniques. This approach was helpful because there are no other suitable standard which can be used to quantify systematically the observed dispersion in the data.

According to the ISO 5725-2, the repeatability standard deviation is a measure of the dispersion of the distribution of independent test results obtained with the same method on identical test items in the same laboratory by the same operator using the same equipment within short intervals of time. The reproducibility standard deviation is a measure of the dispersion of the distribution of test results obtained with the same method on identical test items in different and independent laboratories with different operators using different equipment. Knowing the two standard deviations for all measurement methods it is possible to estimate the precision of the measurement. The two quantities are related by the formula:

$$s_R^2 = s_L^2 + s_r^2 , \tag{1}$$

where $s_L^2$ is the estimate of the between-laboratory variance, $s_r^2$ is the estimate of the repeatability variance, which can be obtained from the mean of the in-laboratory variances and $s_R^2$ is the estimate of the reproducibility variance.

The results were also analysed with the aid of the Mandel's and Cochran's statistical tests described in the ISO 5725-2, in order to evaluate the consistency of the data. With the Mandel's test, the histogram graphs of the parameters *h* and *k*[21,22] are obtained, indicating respectively the between-laboratory and the in-laboratory consistency statistics. In particular, the examination of *h* and *k* plot can indicate those laboratories which exhibit inconsistent results. In addition, the Mandel's test can reveal the presence of two distinct populations of results which reflect the fact that different types of measurement techniques were used by the 14 laboratories. The upper limits values *h* and *k* are generally presented at the 1% and 5% significance level. In this paper 5% significance level was adopted. The ISO 5725-2 also assumes that only small differences exist between laboratories in the in-laboratory variance. However, this is not always the case and to this end the Cochran's test[23] gives an indication of possible exclusion of some laboratory if the value is higher than a critical value (which has been fixed at 5% significance level).

## III. RESULTS

### A. Material homogeneity and anisotropy

Firstly, in order to check the homogeneity of materials each laboratory was asked to measure the density of specimen for all the materials tested. The results are summarised in Figure 3. Combining results from all partners for each material, the relative standard deviation of density (calculated as the percentage ratio between the standard deviation and mean value) was equal to 6-7% for materials A, B and D, 29 % for material C and 17% for material E.

Combining results from all the laboratories for each material, the relative standard deviation for density (calculated as the percentage ratio between the standard deviation and mean value) was equal to 6-7% for materials A, B and D, 29 % for material C and 17% for material E. In order to underline possible anisotropy of tested materials, quasi-static compression tests for determining the storage modulus *E* were carried out by laboratory 3 on cubic shaped specimen in three perpendicular directions (*X* and *Y* in plane, *Z* through thickness) and comparison are depicted in Figure 3 in terms of ratio between the directional Young's moduli, $E_X$ and $E_Y$ and $E_Z$ (the index indicates the direction of measurement).

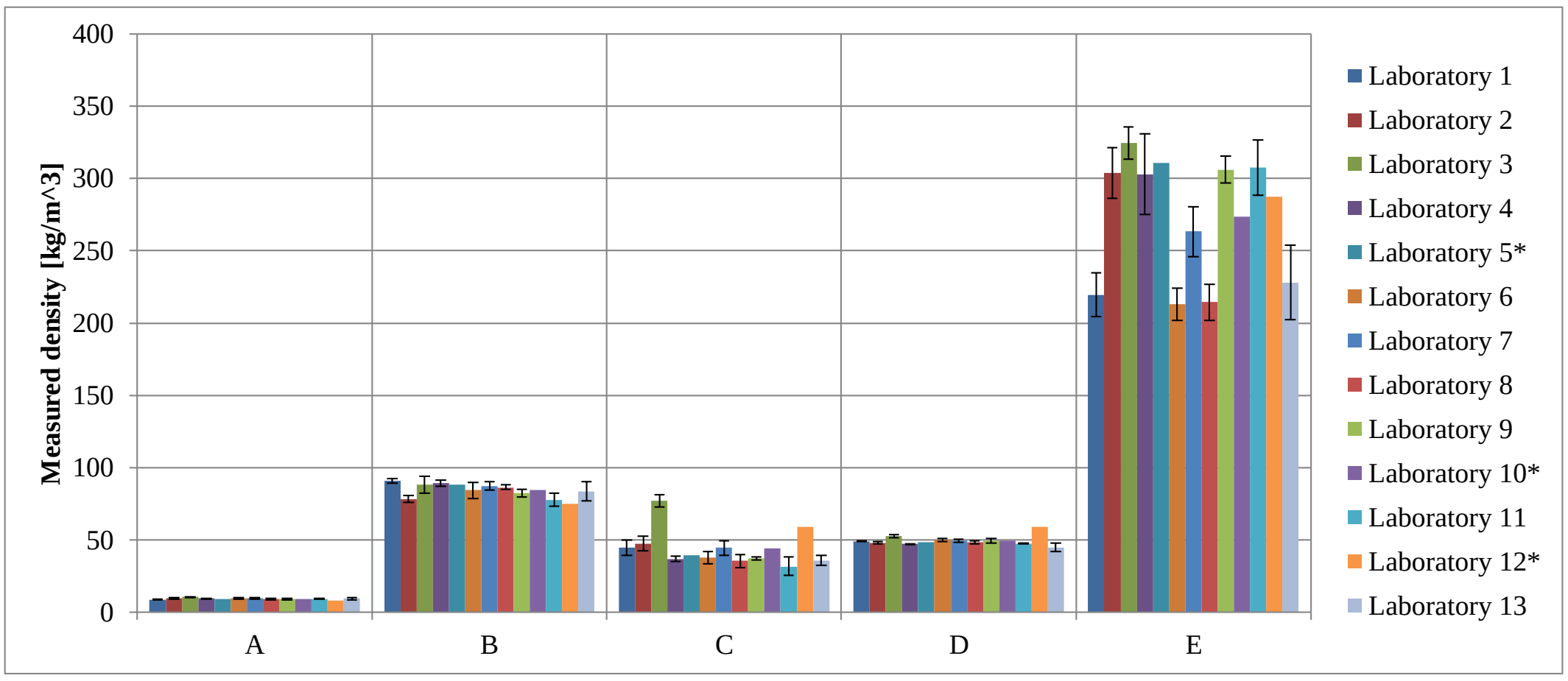

Figure 3 – A comparison of the measured densities (mean value and standard deviation for all specimen). * indicates partners which did not evaluate dispersion of measured density.

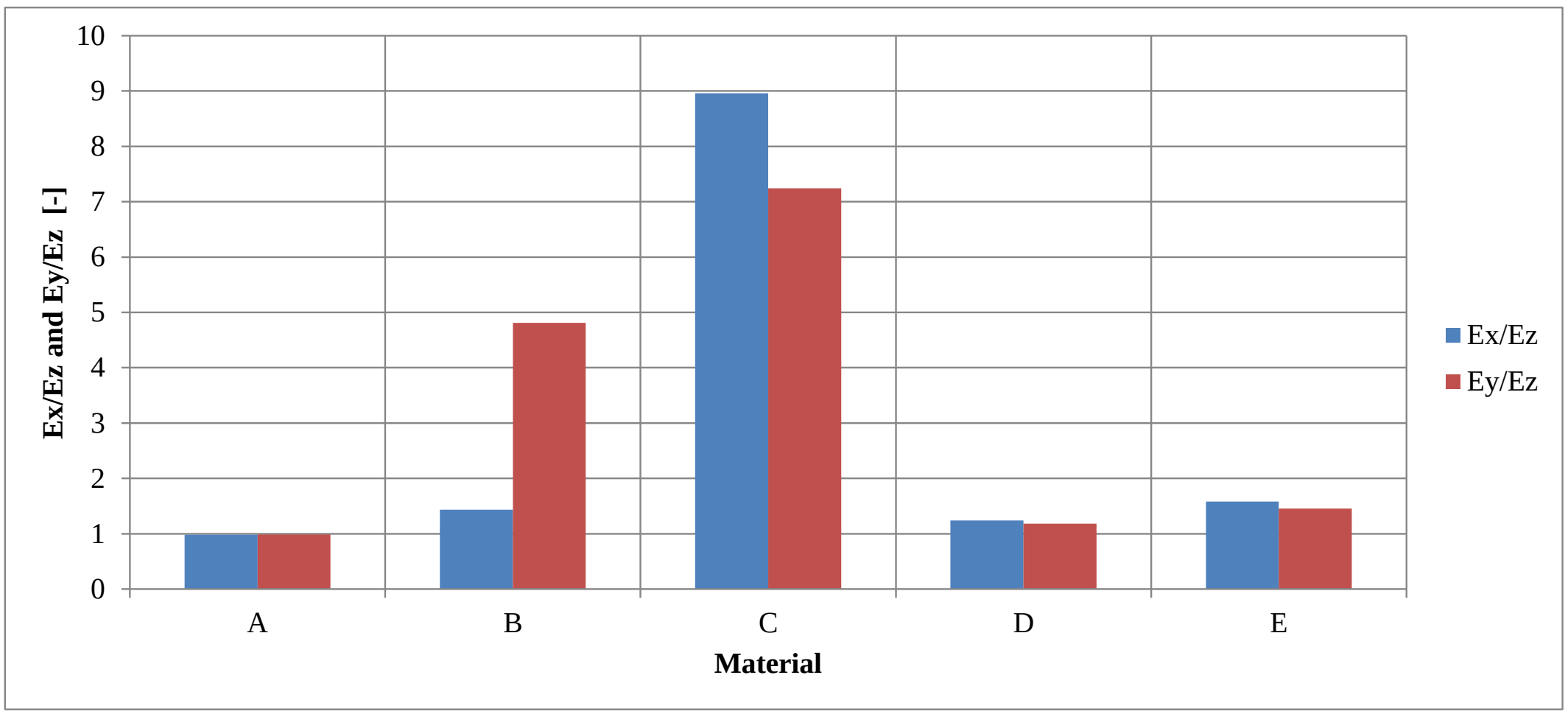

Figure 4 – A comparison of the ratios of in-plane and through-thickness storage modulii for all tested materials carried out by laboratory 3.

From data in Figure 4 it is possible to observe that materials A, D and E are close to being isotropic while there was a significant deviation in the Young's moduli observed for material B in the direction y and for material C in both in-plane directions.

## B. Influence of static preload/compression rate

In order to investigate and quantify the effect of static load, laboratory 3 carried out quasi-static tests using QMA analysis with varying preload in a reduced frequency range (between 30 Hz and 40 Hz) on all the materials varying preload. The results (normalized with respect the value at null load as a ratio for storage modulus and loss factor and as a difference for Poisson's ratio) of these tests are depicted in Figure 5.

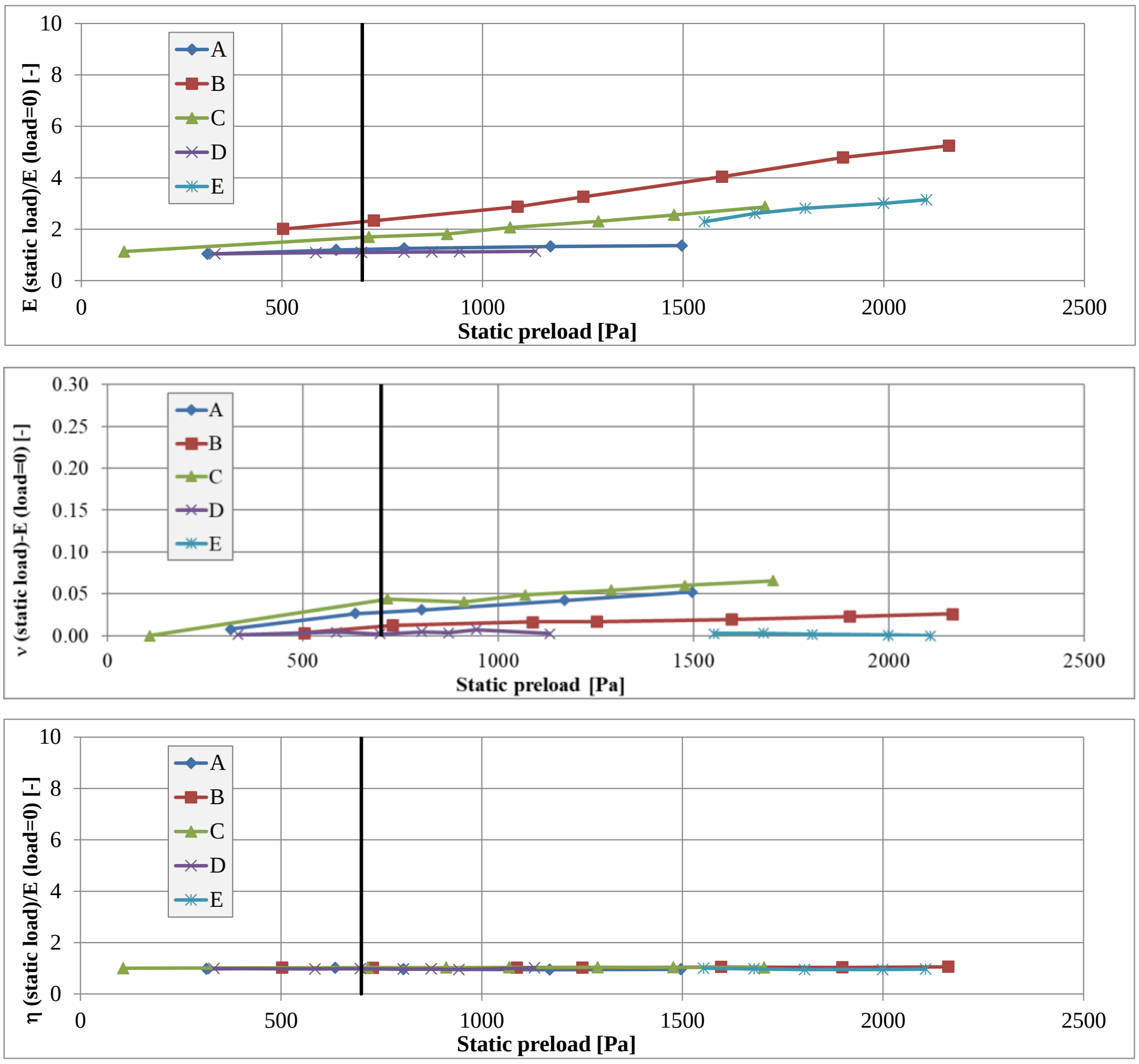

Figure 5 –The dependence of the Young's modulus, Poisson's ratio and loss factor on the static load carried out by laboratory 3.

The results shown in Figure 5 suggest that there was a strong dependence of the Young's modulus on the static preload for materials B, C and E. No significant variation as a function of static preload were observed for the Poisson's ratio and loss factor. Among all participants the maximum static preload was applied by laboratory 4 (~700 Pa) thus a maximum deviation of a factor of 2 for storage modulus and Poisson's ratio was expected according to data depicted Figure 5.

### C. Results of viscoelastic parameters

Figs. 6-10 show comparisons between the storage moduli, Poisson's ratios and loss factors measured by all the 14 laboratories. The values presented in these figures are averaged for all the specimens for each tested material. Figure 11 depicts the overall deviations which were calculated from the difference between minimum and maximum value for each tested material.

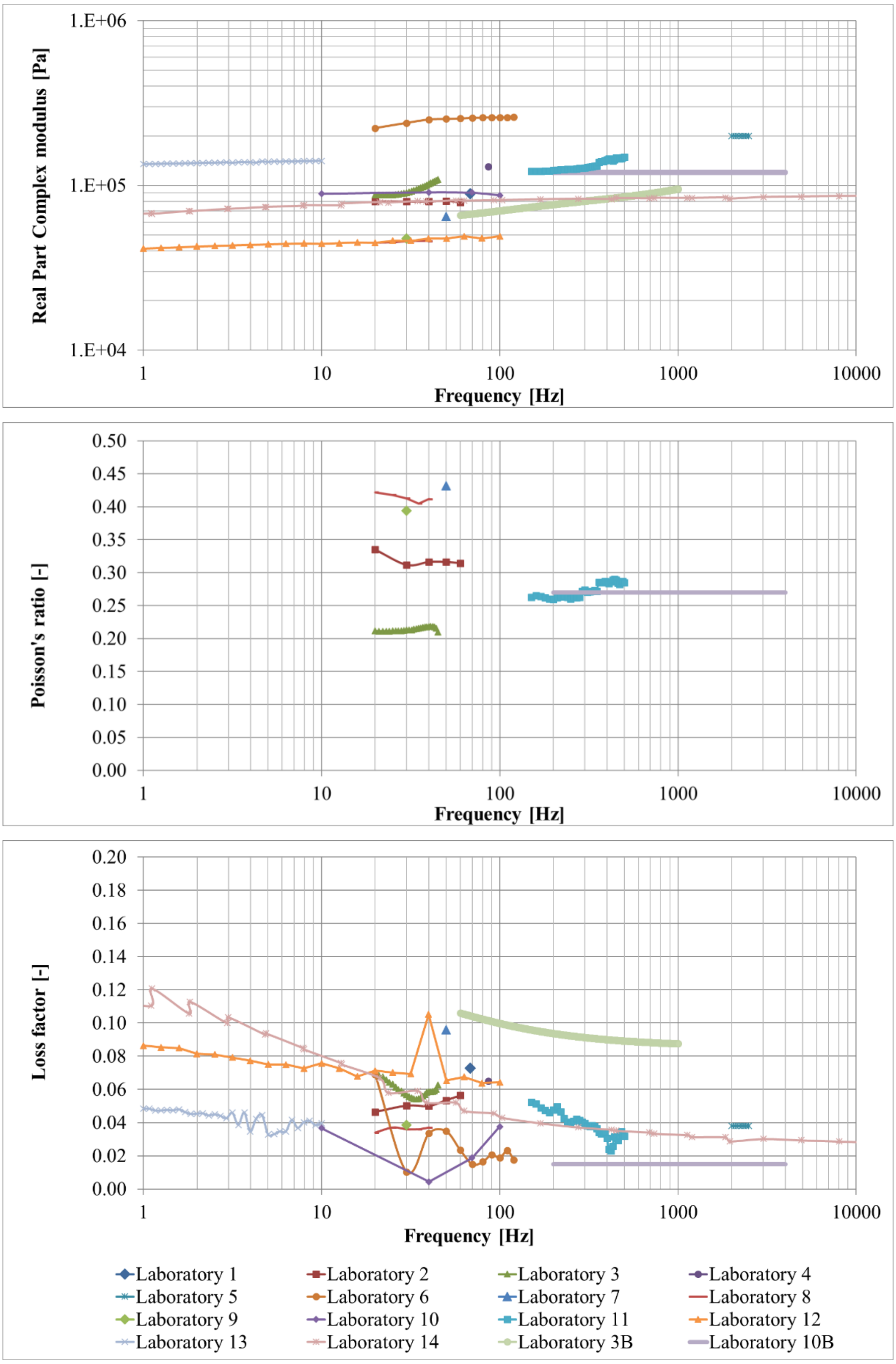

Figure 6 – The Young's moduli (top), Poisson's ratio (middle) and loss factors (bottom) for material A.

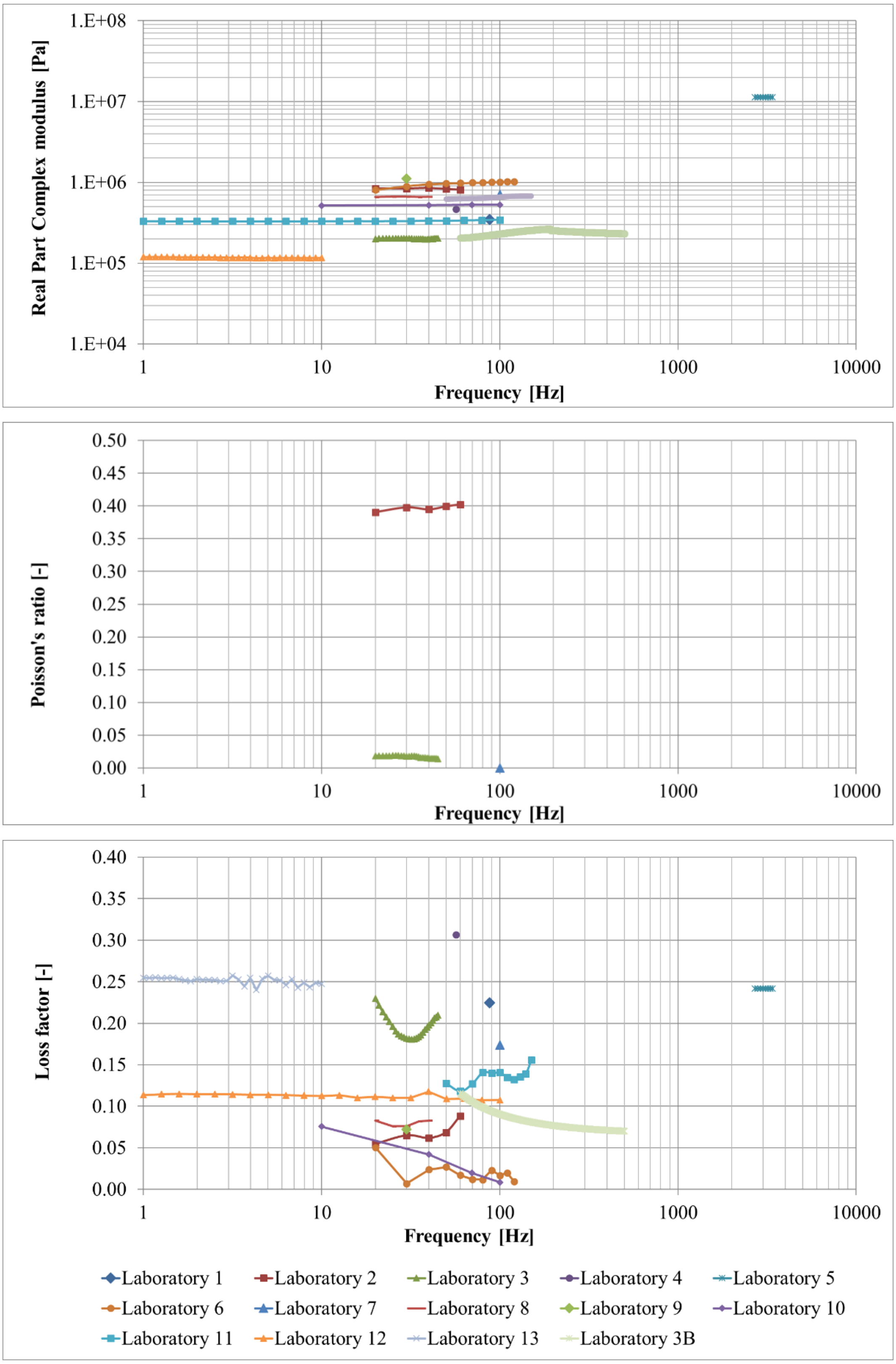

Figure 7 – The Young's moduli (top), Poisson's ratio (middle) and loss factors (bottom) for material B.

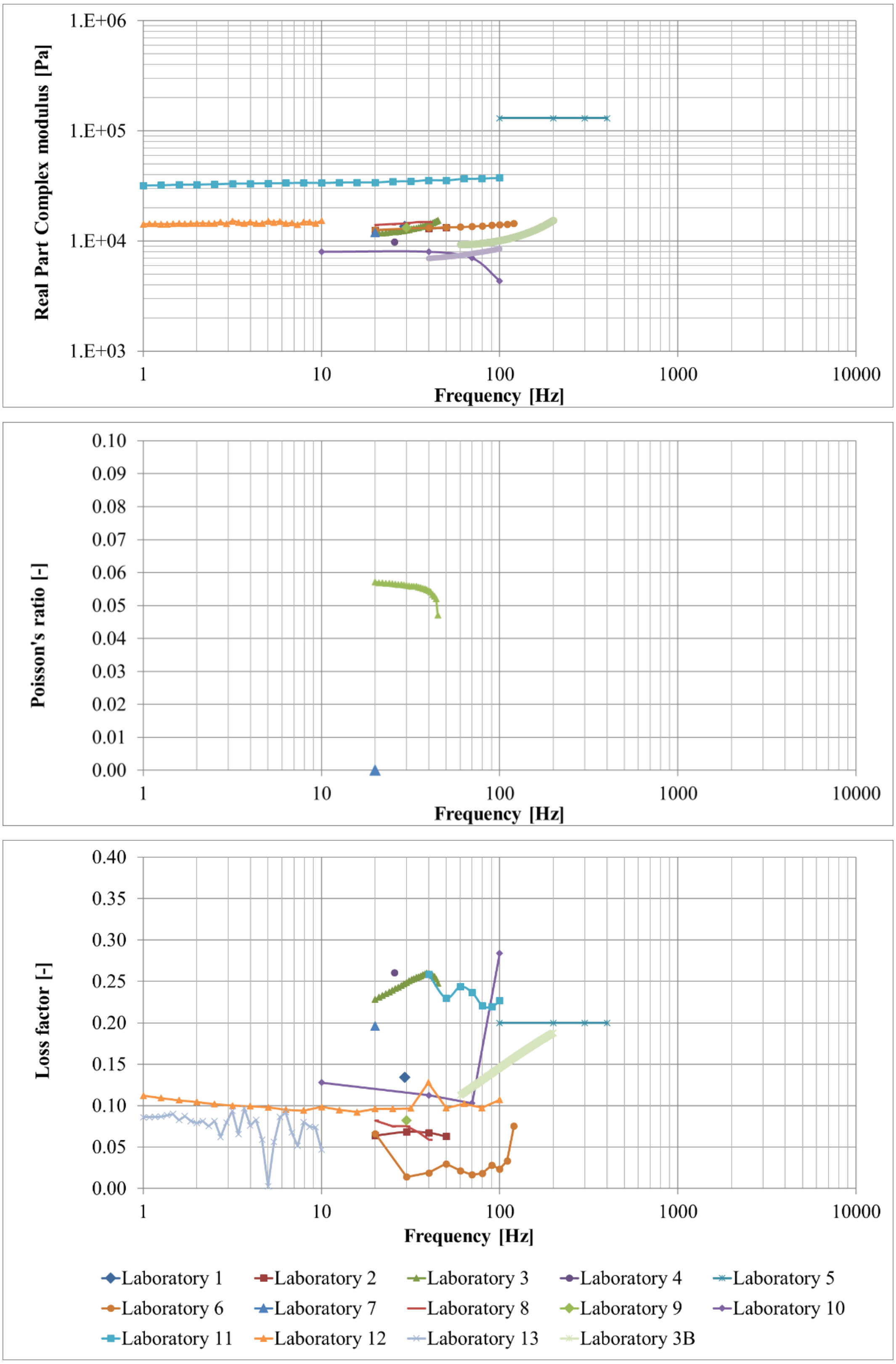

Figure 8 – The Young's moduli (top), Poisson's ratio (middle) and loss factors (bottom) for material C.

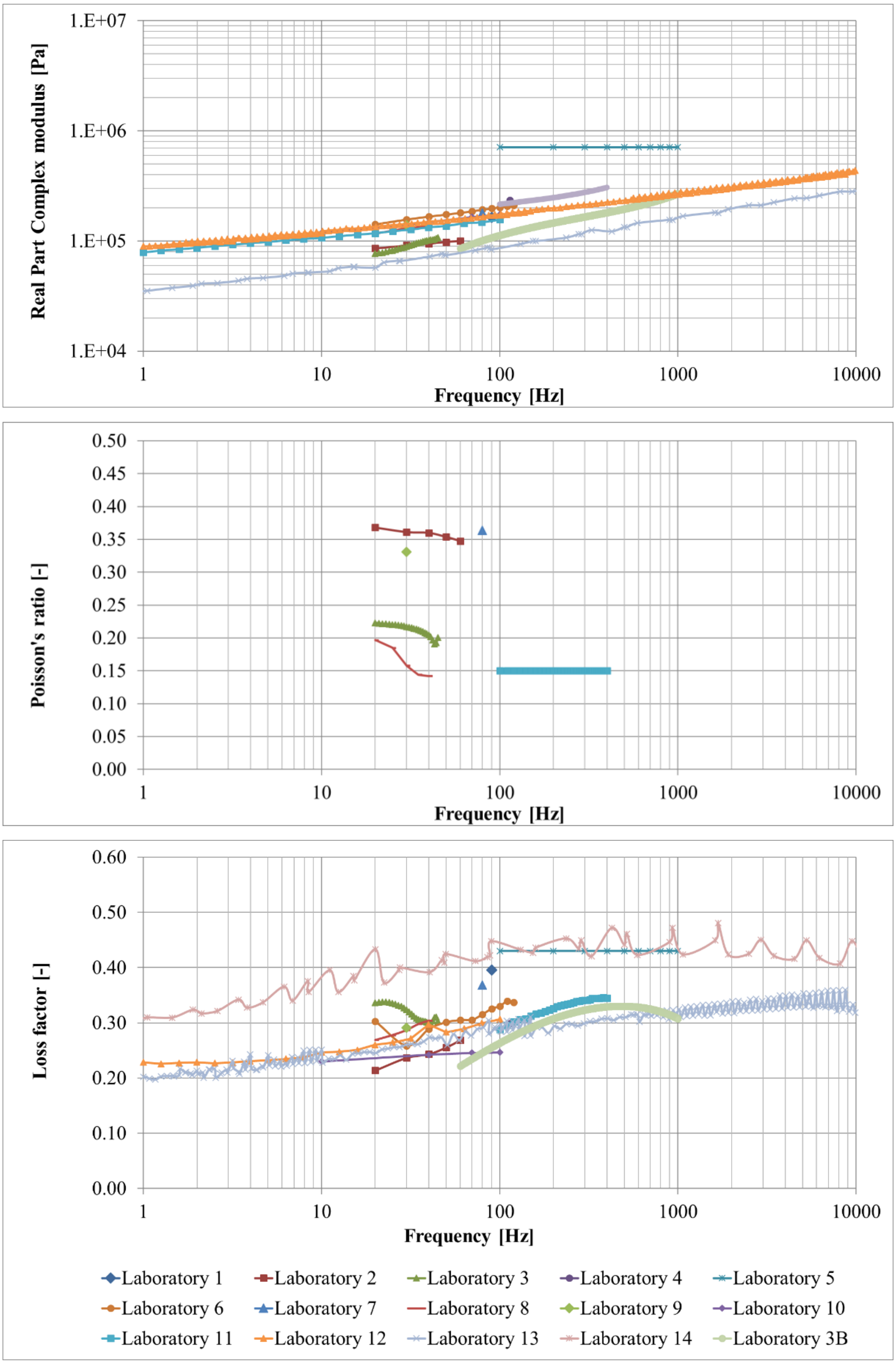

Figure 9 – The Young’s moduli (top), Poisson’s ratio (middle) and loss factors (bottom) for material D.

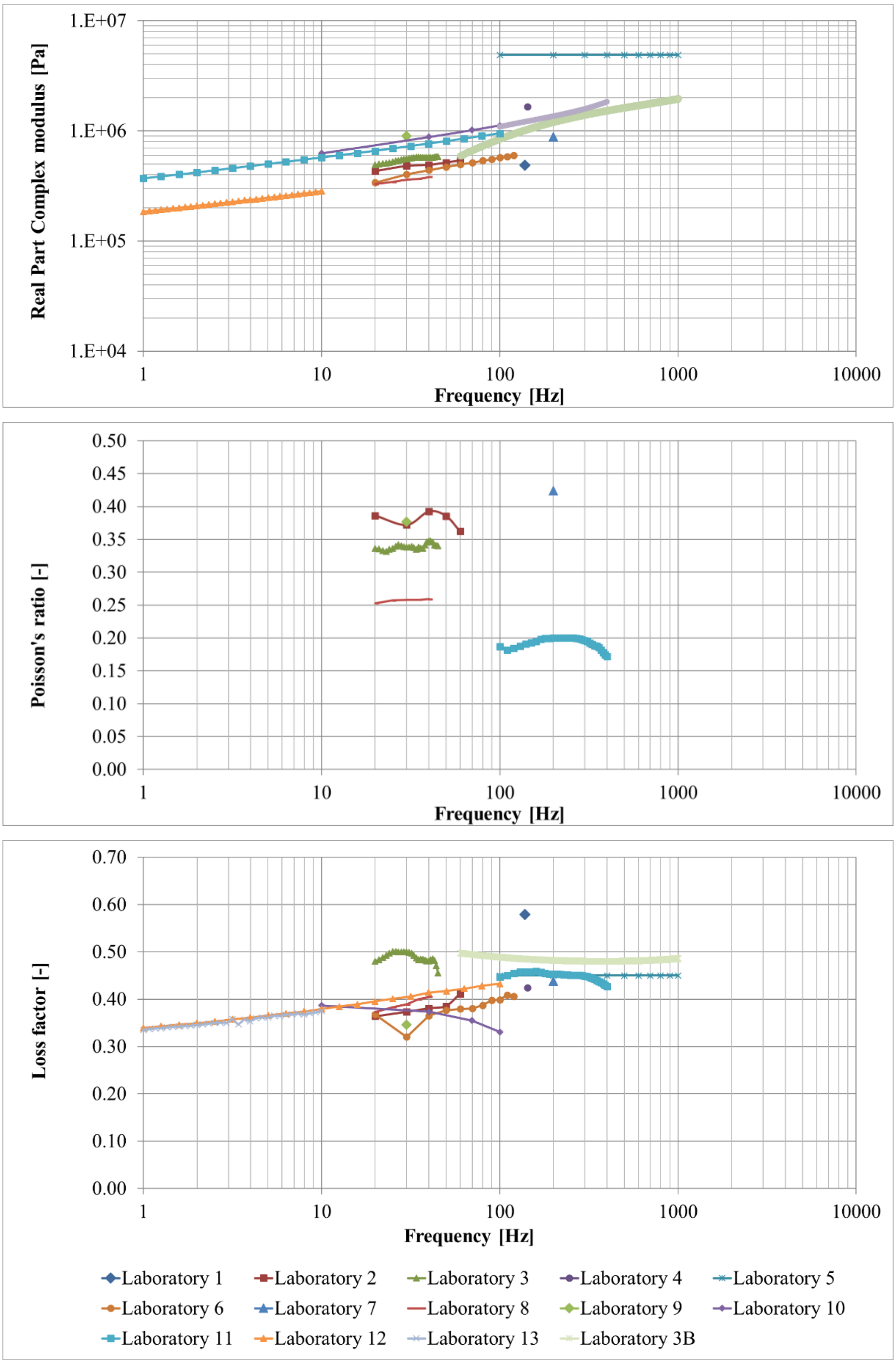

Figure 10 – The Young's moduli (top), Poisson's ratio (middle) and loss factors (bottom) for material E.

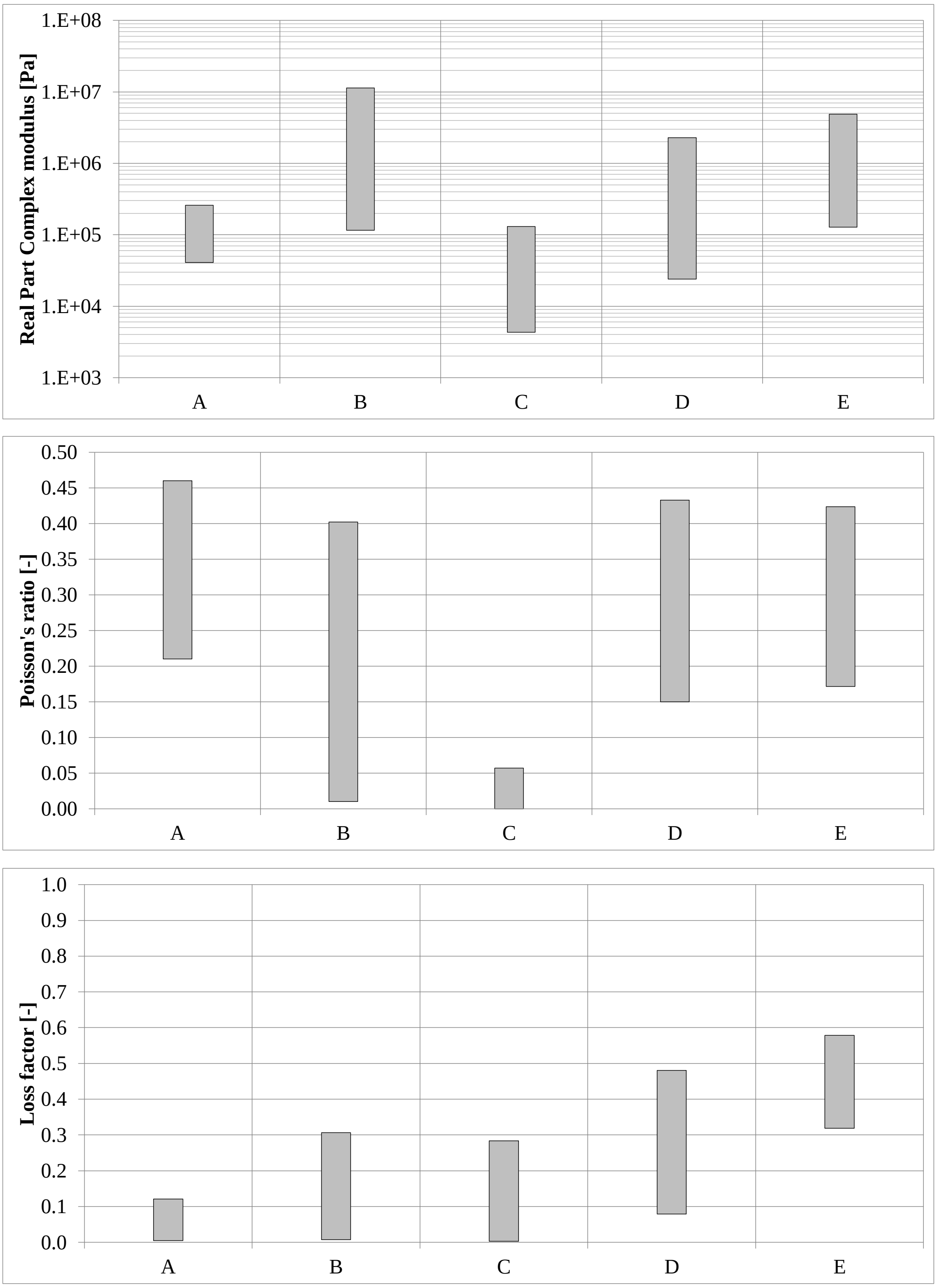

Figure 11 – The overall deviations in the Young's modulus (top), Poisson's ratio (middle) and loss factor (bottom).

Two different reasons could affect the overall standard deviation for the storage modulus, which in some cases reached two orders of magnitude as shown in Figure 11. The first reason was the frequency range for which materials D and E showed a strong viscoelasticity, i.e. a noticeable increase in the storage modulus with frequency. The second reason was skewed data from laboratory 5 who appeared to overestimate the storage modulus significantly and particularly for materials B and C. It is important to remember that the method adopted by laboratory 5 was the unique in terms of testing materials in in-plane direction so that their results can confirm the anisotropy that is typical to fibrous materials as illustrated by data from laboratory 3 in Figure 3. Despite a clear influence of the static preload (Figure 5), this effect did not explain the discrepancies between data provided by laboratories who applied no static load (laboratories 3, 5, 6, 10, 14). Other conditions could have mask the influence of this parameter.

High deviations were observed in the Poisson's ratio although there was a relatively small volume of direct measured data. In particular, laboratory 2 obtained a value of Poisson's ratio which is markedly higher than those obtained by laboratories 3 and 7 for material B. The values obtained for fibrous materials (B, C) were less than 0.05 (except for laboratory 4 and material B). This seems in line with the hypothesis of null Poisson's ratio. The Poisson's ratio for the other materials varied between 0.15 and 0.45, with an average being between 0.30 and 0.35 which was also in line with usual values used for continuous or cellular materials. This rather large uncertainty may be explained by the fact that the value of the Poisson's ratio had an effect almost one order below that of the Young's modulus and that its estimation could be affected by material anisotropy or homogeneity.

The overall deviation in the loss factor was comparable for all materials, except for material D, due to high values measured by laboratories 5 and 14. No clear dependency of viscoelastic properties from static load or compression rate was observed, although this was typical of viscoelastic materials as depicted in Figure 4.

### D. Statistical analysis of the results

As described in Section II C statistical procedures for the analysis according to the ISO 5725-1 and ISO 5725-2 were applied. The laboratories which tested only one sample for each material were excluded from this analysis. All statistical analysis of the measured Young's modulus and loss factor were applied to data obtained at the frequency of excitation of 50 Hz. Data from those laboratories which operated in a different range were extrapolated to 50 Hz. Data from those laboratories which used a single frequency resonant method were added to the statistical analysis without referring to 50 Hz. The first step in the error analysis was to calculate the relative repeatability standard deviation, $s_r$, and the relative reproducibility standard deviation, $s_R$, summarised in Table V.

Table V. Repeatability and reproducibility standard deviation for $E$ and $\eta$.

| *Lab\Test* | A | B | C | D | E |
|---|---|---|---|---|---|
| $s_r$ ($E$) | 46% | 22% | 22% | 5% | 16% |
| $s_R$($E$) | 71% | 57% | 36% | 29% | 34% |
| $s_r$ (η) | 12% | 13% | 9% | 2% | 2% |
| $s_R$(η) | 44% | 69% | 62% | 14% | 17% |

From data in Table V it can be observed that the (in-laboratory) repeatability for storage modulus $E$ was lower than 22% for materials from B to E, while it was equal to 46% for material A. The (in-laboratory) repeatability for loss factor $\eta$ was lower than 13% for all materials.

The reproducibility standard deviation both for storage modulus and loss factor was significant mainly for materials A, B and C. All such results are compared also in terms the average value and standard deviation of $E$ and $\eta$ for each partner and tested material (Figure 12).

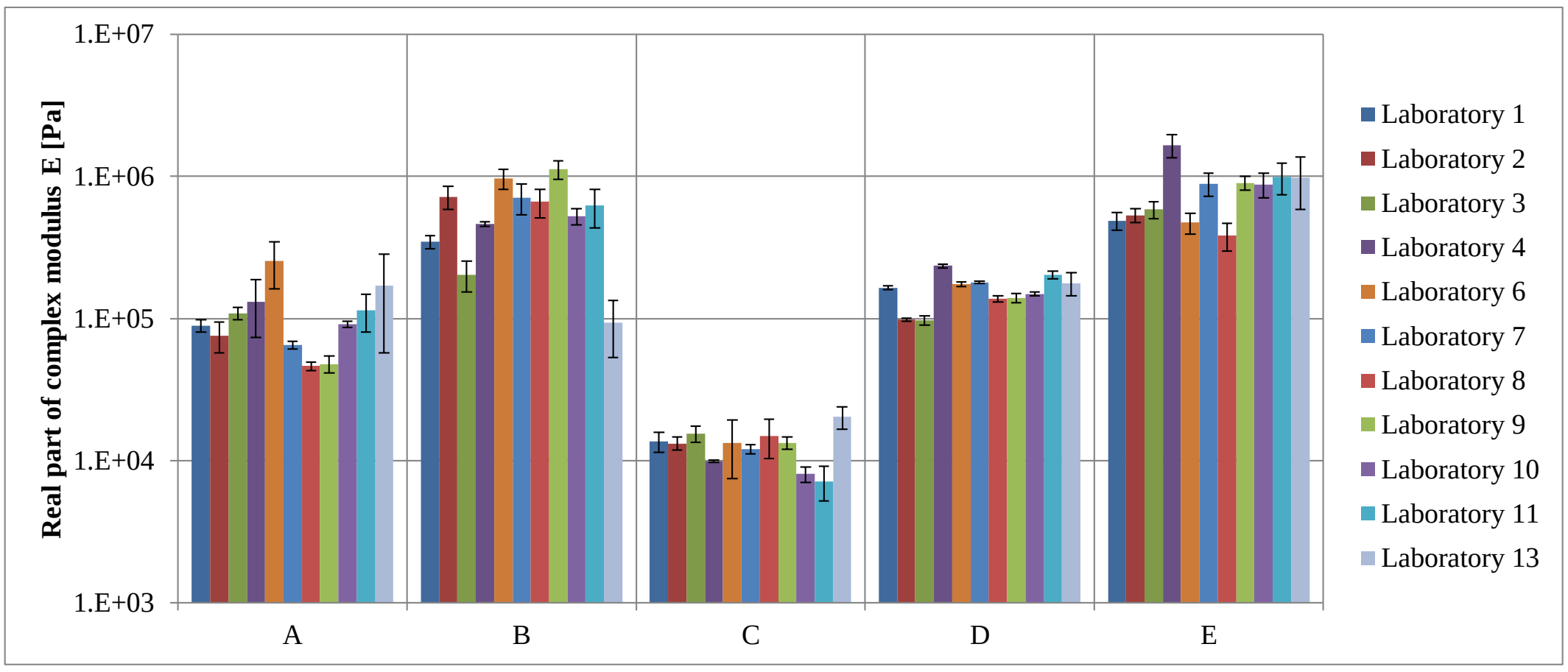

Figure 12 – Comparison in terms of mean value and standard deviation of E for all the laboratories and materials

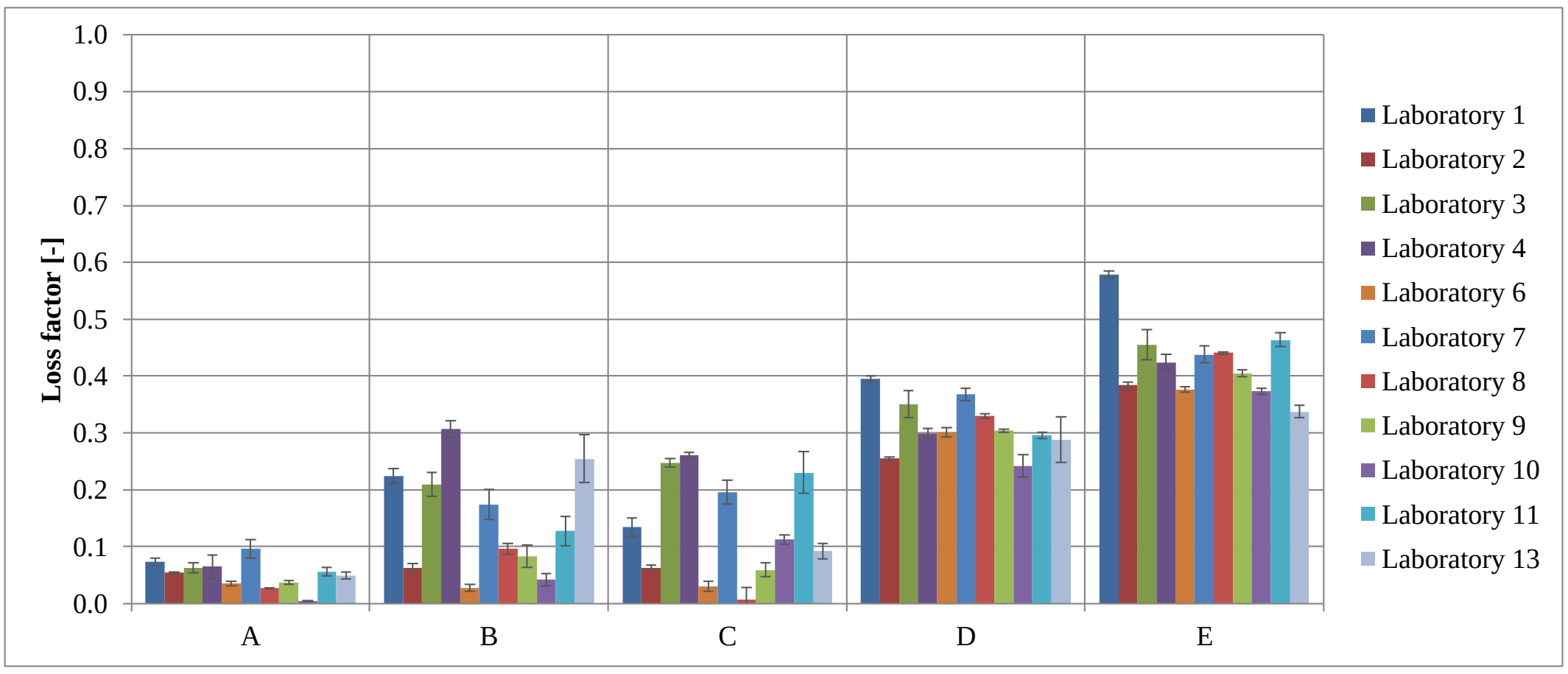

Figure 13 – Comparison in terms of mean value and standard deviation of $\eta$ for all laboratories and materials

The combined results and ISO Standard 5725-2 suggest that laboratories 4, 6, 11 and 13 could strongly affect the repeatability and reproducibility standard deviations shown in Table IV. Such finding relates to the in-laboratory repeatability ($s_r$ and Mandel's *k*-graphs in Figs. 13 and 14) and can also be explained by some degree of inhomogeneity of the tested specimens for each material studied. In fact, for almost all the materials the standard deviation for measured density is higher for above-mentioned partners (see Figure 3). Regarding the between-laboratory results ($s_R$, Mandel's *h*-graph and Cochran's test in Figures 13 and 14), the main differences can be due to a combination of different measurement technique and static load/compression rate initial conditions. The loss factor values were also affected by the type of sample mounting conditions (glue, adhesive tape, sand paper).

Within this context it is not straightforward to separate each contribution since the analysis procedures outlined in the ISO 5725-2 are based on the fact that the same measurement technique was used throughout the inter-laboratory experiment.

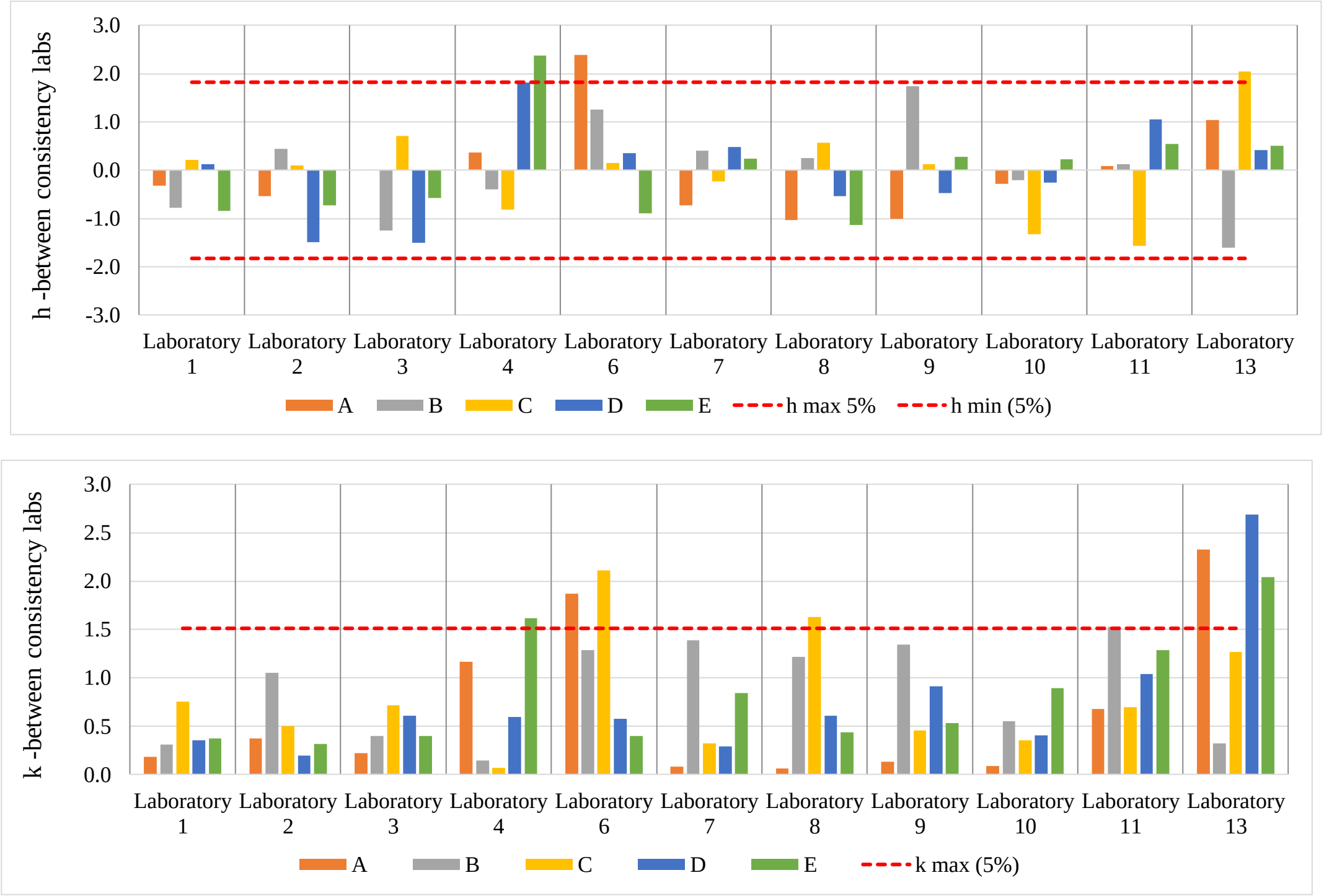

Figure 14 – The values of *h* and *k* Mandel's tests for the storage modulus.

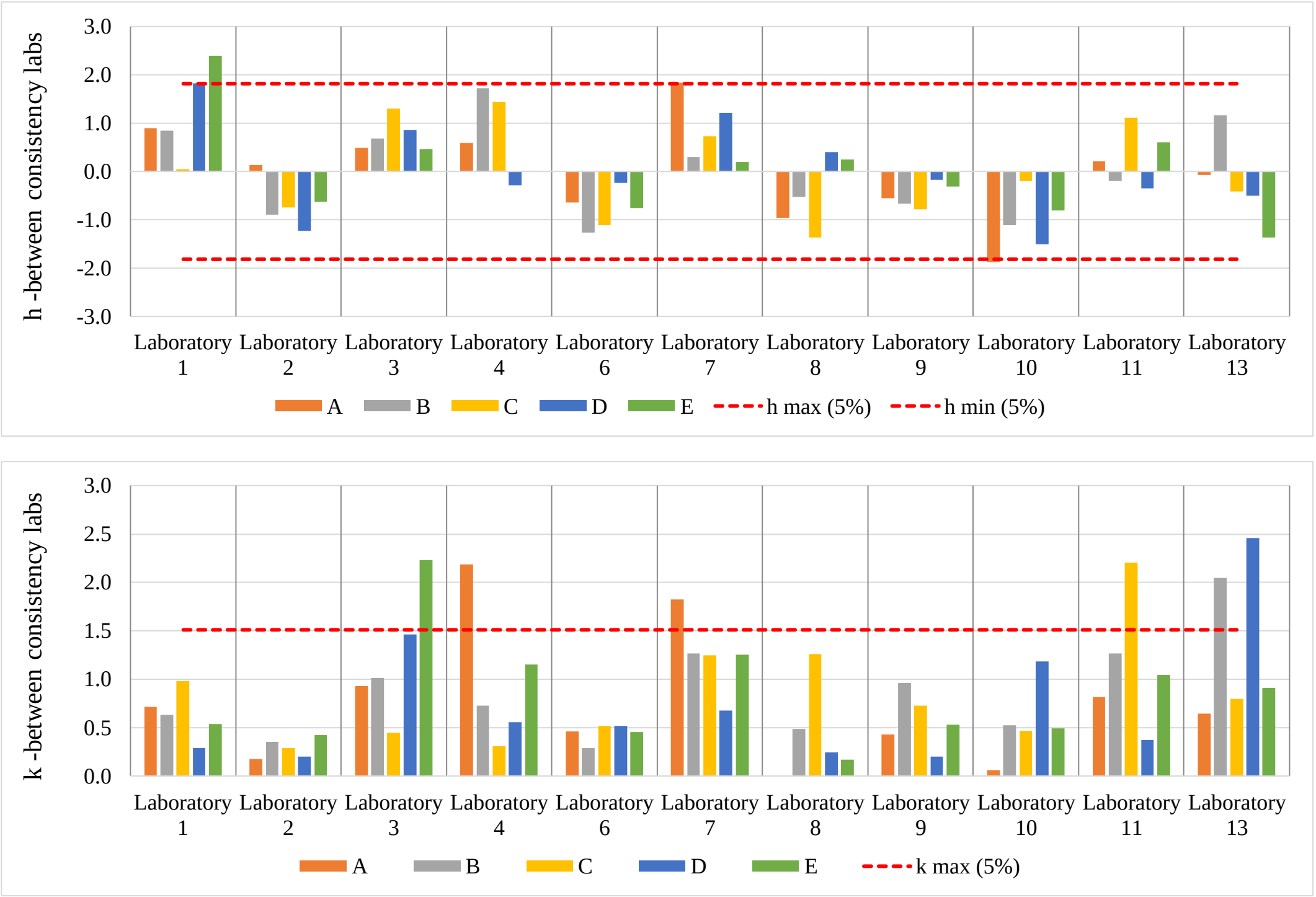

Figure 15 – The values of h and k Mandel's tests for the loss factor.

Table VI. Cochran's test results. (● stands for possible outliers)

| | A | | B | | C | | D | | E | |
|---|---|---|---|---|---|---|---|---|---|---|
| | $E$ | $\eta$ | $E$ | $\eta$ | $E$ | $\eta$ | $E$ | $\eta$ | $E$ | $\eta$ |
| 1 | | | | | | | | | | |
| 2 | | | | | | | | | | |
| 3 | | | | | | | | ● | | ● |
| 4 | ● | ● | | | | | | | ● | |
| 6 | ● | | | | ● | | | | | |
| 7 | | ● | | | | | | | | |
| 8 | | | | | | | | | | |
| 9 | | | | | | | | | | |
| 10 | | | | | | | | ● | | |
| 11 | ● | | | | | ● | | | ● | |
| 13 | ● | | | ● | | | ● | ● | ● | |

## IV. CONCLUSIONS

The inter-laboratory tests on the mechanical properties of 5 types of porous media suggest a poor reproducibility between the 14 participating laboratories. There was a strong dependence of the Young's modulus and loss factor on the static preload and on the test method. An extreme case was the overall deviation in the real part of the Young's modulus for material B (relatively soft glass wool) which varied from the mean by two orders of magnitude. The data on the Young's modulus of material A (relatively stiff reticulated foam) were found to be much more consistent across the independent laboratory tests. The deviation in the Poisson's ratio was found highest for material B, although this parameter was tested by 7 laboratories only. The Poisson's ratio was found to be relatively independent of frequency, but varied considerably between laboratories, e.g. by a factor of 10 for material B. Three possible reasons for these results are: (i) a strong frequency and temperature dependence of the viscoelastic properties; (ii) the presence of significant outliers in the results from some laboratories (e.g. laboratory 5); (iii) material anisotropy particularly in the case of glass wool; (iv) the inhomogeneity of the materials.

The deviation in the loss factor data was found comparable for all the materials except material D (close cell polyurethane foam). Laboratories 5 and 14 overestimated heavily the value of the loss factor for material D. Laboratory 5 used the Lamb wave method and laboratory 14 used the dynamic mechanical analysis method with the subsequent time-temperature superposition to extend the frequency range. These methods involved different solicitations of the material to which the loss factor could be sensitive.

The results of the error analysis carried out in accordance with the ISO 5725 Parts 1 and 2 suggest that the maximum relative reproducibility standard deviation in the measurement of the Young's modulus was 71% for material A. The maximum relative reproducibility standard deviation in the measurement of the Poisson's ratio was 62% for material C (felt). The reproducibility standard deviation was also significant for material B.

These findings suggest that there is an obvious need for harmonisation of the procedures to measure the complex Young's modulus and Poisson's ratio of porous media. There is no agreed guidance on the preparation and installation of the samples during the test, no instrument calibration procedures or procedures for periodic verification of the instruments and no guide to verify that the hypotheses made for a given test are a valid posteriori. There is no guidance on the number of samples to be measured for the characterisation of a material and the acceptability of a certain standard deviation on the tests conducted is not agreed. It is recommended that a steering group is setup to propose a new international standard for testing the mechanical properties of porous media.

**AKNOWLEDGEMENTS**

Authors would like to acknowledge companies Cofermetal, Isover, Adler Pelzer and L'Isolante K-FLEX for providing materials used for the interlaboratory test. This article is based upon work from COST Action DENORMS CA 15125, supported by COST (European Cooperation in Science and Technology).

Figure 1 – Tested materials

Figure 2 – Basic measurement setups for: a) and b) quasi-static uniaxial compression methods, c) and d) resonant methods, e) dynamic torsional method, f) Lamb wave propagation method, g) Surface acoustic wave method, h) transfer function/transfer matrix method. 1-sample; 2-accelerometer; 3-force transducer;4-torque transducer; 5-angular displacement transducer; (6) laser vibrometer.

Figure 3 – A comparison of the measured densities (mean value and standard deviation for all specimen). * indicates partners which did not evaluate dispersion of measured density.

Figure 4 – A comparison of the ratios of in-plane and through-thickness storage moduli for all tested materials carried out by Partner 3.

Figure 5 –The dependence of the Young's modulus (top), Poisson's ratio (middle) and loss factor (bottom) on the static load carried out by Partner 3.

Figure 6 – The Young's moduli (top), Poisson's ratio (middle) and loss factors (bottom) for material A.

Figure 7 – The Young's moduli (top), Poisson's ratio (middle) and loss factors (bottom) for material B.

Figure 8 – The Young's moduli (top), Poisson's ratio (middle) and loss factors (bottom) for material C.

Figure 9 – The Young's moduli (top), Poisson's ratio (middle) and loss factors (bottom) for material D.

Figure 10 – The Young's moduli (top), Poisson's ratio (middle) and loss factors (bottom) for material E.

Figure 11 – The overall deviations in the Young's modulus (top), Poisson's ratio (middle) and loss factor (bottom).

Figure 12 – Comparison in terms of mean value and standard deviation of $E$ for all the laboratories and materials.

Figure 13 – Comparison in terms of mean value and standard deviation of $\eta$ for all laboratories and materials

Figure 14 – The values of h and k Mandel's tests for the storage modulus.

Figure 15 – The values of h and k Mandel's tests for the loss factor.